\documentclass[epj,referee]{svjour}
\usepackage{graphics}

\begin{document}

\title{Buckyball Quantum Computer: Realization of a Quantum Gate}

\author{Maria Silvia Garelli \and Feodor V Kusmartsev}
\institute{Department of Physics, Loughborough University, LE11 3TU, UK \email {M.S.Garelli@lboro.ac.uk }}

\date{Received: date / Revised version: date}
\abstract{
We have studied a system composed by two endohedral fullerene
molecules. We have found that this system can be used as good candidate for the realization of Quantum Gates. Each of these molecules encapsules an atom carrying a
spin, therefore they interact through the spin dipole interaction.
We show that a phase gate can be realized if we apply static and time dependent magnetic fields on each encased spin. We have
evaluated the operational time of a $\pi$-phase gate, which is of the order of $ns$. We made a comparison between the theoretical estimation of the gate time and the
experimental decoherence time for each spin. The comparison shows
that the spin relaxation time is much larger than the $\pi$-gate
operational time. Therefore, this indicates that, during the
decoherence time, it is possible to perform some thousands of
quantum computational operations. Moreover, through the study of concurrence, we get very good results for the entanglement degree of the two-qubit system. This finding opens a new avenue
for the realization of Quantum Computers.
} 
\maketitle

\section{Introduction}
During recent years there is a strong progress in modeling physical
realizations of a quantum computer. Many quantum physical systems have been investigated for the realization of quantum gates. The most remarkable studies were related to systems associated to Quantum Optics Ion Traps, to Quantum
Electrodynamics in Optical Cavities and to Nuclear Magnetic Resonance. All these experiments are aimed to realize a quantum gate. The first type of experiments is based on trapping ions in electromagnetic traps, where the ions, which encode the qubit in the charge degrees of freedom, are subjected to the mutual electrostatic interaction and to a state selective displacement generated by an external state dependent force \cite{Cirac,Steane,Sasura,Calarco}.
Cavity quantum electrodynamics (QED) techniques are based on the coherent interaction of a qubit, generally represented by an atom or semiconductor dot system, with a single mode or a few modes of the electromagnetic field inside a cavity. Depending on the particular system, the qubit can be represented by the polarization states of a single photon or by two excited states of an atom. Although cavity QED experiments are very promising, they have been accomplished for few qubits \cite{Pellizzari,van,Rauschenbeutel,Duan}. In the third experiment, nuclear spins represent qubits. These spins can be manipulated using nuclear magnetic resonance techniques, and through the study of the quantum behavior of spins, quantum operations are realized. However, the number of spins which can be collected in a system is very limited, and this forbids the building up of a scalable quantum computer \cite{Gershenfeld,Schmidt,Leibfried,Nielsen}.  
From the study of such systems, we learn that the decoherence phenomenon is the main issue which
prevents the realization of quantum gates. Here we will focus on
a physical systems, which will be able to produce a realistic quantum gate.
The basic elements of our system are fullerene molecules
with encapsulated atoms or ions, which are called
\emph{buckyballs} or \emph{endohedral fullerenes}. Each of the
trapped atoms carries a spin. This spin, associated with
electronic degrees of freedom, encodes the qubit. It has been
shown \cite{Greer}, that these endohedral systems
provide a long lifetime for the trapped spins and that the
fullerene molecules represent a good sheltering environment for the very
sensible spins trapped inside. These endohedral systems are
typically characterized by two relaxation times. The first is
$T_1$, which is due to the interactions between a spin and the
surrounding environment. The second one is $T_2$ and it is due to
the dipolar interaction between the qubit encoding spin and the
surrounding endohedral spins randomly distributed in the sample. While
$T_1$ is dependent on temperature, $T_2$ is practically
independent of it. The experimental measure of the two relaxation
times shows that $T_1$ increases with decreasing temperature from
about $100\mu s$ at $T=300 K$ to several seconds below $T=5K$, and
that the value of the other relaxation time, $T_2$, remains constant, that is  $T_2\simeq 20\mu s$ \cite{Knorr1,Knorr2}. In
comparison with $T_2$ the value of $T_1$ is very large, therefore
the system decoherence is determined by the spin-spin relaxation
processes. It is supposed that the value of $T_2$ can be
increased, if it will be possible to design a careful experimental
architecture, which could screen the interaction of the spins with the
surrounding magnetic moments. It should be possible to reduce the relaxation time of the system due to the random spin-spin interactions, if we
consider a system composed by arrays of endohedrals encapsulated
in a nanotube \cite{Khlobystov}, this system is also called as \emph{peapod}, or considering buckyballs embedded on a
substrate.
These should be reliable systems for the realization of
quantum gates. In such architectures the decoherence time for each
encapsulated spin should be longer. 

 Quantum computing through the study of doped fullerene systems has been investigated in many works \cite{Harneit,Harneit1,Feng,Suter,Twam}. Although we have followed many ideas suggested in these previous papers, we consider a different approach for the realization of quantum gates. 
 
 Our study is focused on a
system composed by two buckyballs. Our aim is the realization of a
quantum \emph{$\pi$-gate}, which is a generalization of the
\emph{phase gate}, this will be treated in Sec. \ref{phasegate}. To perform the $\pi$-gate, we need to know the
time evolution of the coefficients of the standard computational
basis states over which we expand the wave function of our system.
The two particle phases are evaluated through the numerical
solution of the Schr{\"o}dinger equation, see Secs. \ref{ind}-\ref{dip}. We have used two approaches: a time
independent Hamiltonian, see 
Sec. \ref{ind}, and a time
dependent one, see Sec. \ref{dip}. The main result of our study
is the gate time, that is the time required by the system in order
to perform the $\pi$-gate. The values obtained are around
$\tau\simeq1\times10^{-8}s$, which is a few orders smaller than
the shortest relaxation time, $T_2$. From the comparison of the
gate time, $\tau$, to the relaxation time, $T_2$, we get that it
is theoretically possible to realize some thousands of basic gate
operations before the system decoheres. We have also checked the
reliability of our gate through the analysis of the
\emph{concurrence} of the two-qubit state, see Sec. \ref{concurr}. The best value for the concurrence is obtained in
the case of a time dependent Hamiltonian, while the gate time is
nearly the same in both cases.

\section{Physical Features of the System}
The system under consideration is composed by two interacting buckyballs.
Several experimental and theoretical studies on buckyballs \cite{Greer,Harneit,Heath,Shinohara,Saunders,Weid}, show that many different types of atoms can be encased in fullerenes molecules. However, in most of the studied endohedral \\
fullerenes, there is a charge transfer from the encapsulated atom to the fullerene cage, with a resulting considerable alteration of the electronic properties of the cage. This is not the case for group V encased atoms. These atoms reside just at
the center of the fullerene molecule, therefore there is no hybrididazion of the electron cloud of the encased atom and there is no Coulomb interaction with the fullerene
cage. In particular, the most promising endohedral molecule should
be the $N@C_{60}$, which is characterized by many interesting chemical-physical properties.
Following Refs. \cite{Greer,Harneit,Weid}
, experiments and theoretical calculations suggest that there is a repulsive
exchange interaction between the fullerene and the electronic
cloud of the encapsulated atom. The electrons in the cloud of the encased nitrogen are tighter bound than in a free nitrogen atom, which allow the encased nitrogen to be less reactive even at room temperature. These results,
together with the location of the nitrogen atom in the central
site, suggest that in $N@C_{60}$ the nitrogen can be considered as
an independent particle, with all the properties of the free atom.
Since any charge interaction is screened, the fullerene cage does not take any part in the interaction process and it can be considered just as a trap for the nitrogen atom. 
Therefore, the only
physical quantity of interest is the spin of the trapped particle.
A nitrogen atom can be effectively described as a
$\frac{3}{2}$-spin particle. This spin is associated with the
electronic degrees of freedom. Taking into account also the nuclear spin, which is $\frac{1}{2}$ for the $N@C_{60}$, the number of relevant degrees of freedom will be not increased \cite{Meher}. We will consider a more simple
model assuming that the encased atoms are described as
$\frac{1}{2}$-spin particles. In absence of any mutual interaction
and without any applied magnetic field, the energy levels associated with
these spin particles are degenerate. If we apply a static magnetic
field, this degeneracy is lifted. As a result, due to the Zeeman
effect, a two level system arises for each $\frac{1}{2}$-spin
particle. Each of these two
levels encodes the qubit. The spin-up component,
$m_s=+\frac{1}{2}$, encodes the computational basis state
$\mid1\rangle$, and the spin-down component, $m_s=-\frac{1}{2}$,
represents the state $\mid0\rangle$.
\section{Gate Operation: The Phase Gate}\label{phasegate}
Quantum computers operate with the use of \emph{Quantum Gates}.
Quantum gates are defined as fundamental quantum computational
operations. They are presented as unitary transformations, which
act on the quantum states, which describe the qubits. 
Therefore a quantum computer must operate with the use of many
quantum gates. The simplest gates are the single-qubit gates.
Since our system is composed by two qubits, we will consider a
two-qubit quantum gate. One of the most important quantum gates is
the \emph{Universal Two-Qubit Quantum Gate} \cite{Nielsen}, which
is called the CNOT-gate. The CNOT operation is defined by the
following four by four unitary matrix
\begin{equation}\label{cnot}
U_{CNOT}=
\left(%
\begin{array}{cccc}
1 & 0 & 0 & 0 \\
0 & 1 & 0 & 0 \\
0 & 0 & 0 & 1 \\
0 & 0 & 1 & 0\\
\end{array}%
\right),
\end{equation}
and its action over the computational basis states reads:
\begin{eqnarray}
\mid 00\rangle &\rightarrow &\mid 00\rangle ;\\
\mid 01\rangle &\rightarrow &\mid 01\rangle ;\\
\mid 10\rangle &\rightarrow &\mid 11\rangle ;\\
\mid 11\rangle &\rightarrow &\mid 10\rangle .
\end{eqnarray}
The CNOT gate is given by the composition of a single-qubit
Hadamard gate followed by a two-qubit $\pi$-gate, finally followed
by another single-qubit Hadamard gate. The representation of the
Hadamard gate in the Bloch sphere is a $\frac{\pi}{2}$ rotation
about the $y$ axis, followed by a reflection of the $x-y$ plane.
In this paper we will focus on the realization of the two-qubit
$\pi$-gate. It is a particular choice of the general \emph{phase
gate}, represented by the following matrix
\begin{equation}
G=
\left(%
\begin{array}{cccc}
  1 & 0 & 0 & 0 \\
  0 & 1 & 0 & 0 \\
  0 & 0 & 1 & 0 \\
  0 & 0 & 0 & e^{\imath\vartheta} \\
\end{array}%
\right),
\end{equation}
and its action on the computational basis states is the following:
 \begin{eqnarray}\label{phasegate1}
|00\rangle &\rightarrow& |00\rangle  \\
|01\rangle&\rightarrow& |01\rangle  \\
|10\rangle&\rightarrow& |10\rangle \\
\label{phasegate4}
 |11\rangle&\rightarrow&e^{\imath
\vartheta}|11\rangle.
\end{eqnarray}
 When $\vartheta=\pm\pi$, the resulting quantum gate is called a $\pi
 -$\emph{gate}. In general, the time evolution of the four states of
the standard computational basis can be described  as follows:
\begin{eqnarray}\label{phase1}
|00\rangle &\rightarrow&e^{i
\phi_{00}} |00\rangle  \\
|01\rangle&\rightarrow&e^{i
\phi_{01}} |01\rangle  \\
|10\rangle&\rightarrow&e^{i
\phi_{10}} |10\rangle \\
\label{phase4} |11\rangle&\rightarrow&e^{i \phi_{11}}\mid
11\rangle.
\end{eqnarray}
In order to obtain the action of the ideal quantum phase gate,
equations (\ref{phasegate1}-\ref{phasegate4}), see Ref. \cite{Calarco}
, we have to apply the following local operator:
\begin{equation}
\hat S=\hat S_1\otimes \hat S_2,
\end{equation}
where
\begin{eqnarray}
\hat S_1=\mid 0\rangle_1\langle 0\mid e^{\imath s^0_1}+\mid
1\rangle_1\langle 1 \mid e^{\imath s^1_1}\\
\hat S_2=\mid 0\rangle_2\langle 0\mid e^{\imath s^0_2}+\mid
1\rangle_2\langle 1 \mid e^{\imath s^1_2}
\end{eqnarray}
and the phases $s^j_i$ are defined as follows::
\begin{eqnarray}
s^0_1&=&-\phi_{00}/2\\
s^1_1&=&-\phi_{10}+\phi_{00}/2\\
s^0_2&=&-\phi_{00}/2 \\
s^1_2&=&-\phi_{01}+\phi_{00}/2,
\end{eqnarray}
After a straightforward  calculation we obtain the desirable
phase:
\begin{equation}\label{vartheta}
\vartheta=\phi_{11}-\phi_{10}-\phi_{01}+\phi_{00}.
\end{equation}
In our system, in order to realize a $\pi$-gate, we need to know
the time evolution of the wave function. The time evolved
wave function, expanded on the standard computational basis, is
given by the following equation:
\begin{eqnarray}\label{wave}
\mid \psi(t)\rangle&=c_1(t)\mid 00\rangle+c_2(t)\mid
01\rangle\\\nonumber
&+c_3(t)\mid 10\rangle+c_4(t)\mid 11\rangle.
\end{eqnarray}
Each coefficient $c_i(t)$, $i=1,..4$, is a complex number, whose
phase, arranged as in equation (\ref{vartheta}), is used for the
realization of the $\pi$-gate.
\section{Concurrence}\label{concurr}
When we consider a $\frac{1}{2}$-spin particle as the encoding
system for the qubit, it may incur to a \emph{spin-flip} process.
This phenomenon consists in the swapping between the spin-up and
spin-down components
\begin{eqnarray}
\mid0\rangle & \rightarrow & \mid1\rangle,\\
\mid1\rangle & \rightarrow & \mid0\rangle.
\end{eqnarray}
If we consider the two-qubit state, known as
\emph{EPR pair},
\begin{equation}
\frac{\mid00\rangle+\mid11\rangle}{\sqrt{2}},
\end{equation}
we can see that it is unaffected by the spin-flip of both qubits.
This state, for this feature, is called maximally
\emph{entangled}. Therefore, we can define the \emph{entanglement}
as the property of quantum states, which shows if the state is
good for carrying quantum information. The most entangled a
quantum state is, the most reliable it is for transferring quantum
information. In our study we have considered the \emph{concurrence}, see Ref. \cite{Wootters}, as a measure of the entanglement of the state
describing the two-qubit system. A \emph{pure} state of two
particles quantum system is called entangled if it cannot be
factorisable, that is it cannot be written as the direct product
of the states describing each particle. A \emph{mixed} state if it
cannot be represented as a mixture of factorisable pure states. In
this Section we will refer to the \emph{entanglement of
formation}, which quantifies the resources needed for the creation of an entangled state. For a complete treatment about the entanglement of formation of pure and mixed states see Refs. \cite{Bennet,Hill}.
The entanglement of formation of a quantum state can be evaluated through the concurrence \cite{Wootters}. Since the state describing our system is a pure state, the degree of entanglement of our system can be quantified through the definition of  the concurrence for a pure state \cite{Wootters}, which is defined by
\begin{equation}\label{concurrence}
C(\psi)=\mid\langle\psi\mid\tilde\psi\rangle\mid,
\end{equation}
where $\mid \tilde\psi\rangle$ is the spin-flipped state of system. The spin-flip
transformation, which for a $\frac{1}{2}$-spin particle is the
standard time reversal transformation \cite{sakurai}, is defined as follows
\begin{equation}\label{spflip}
\mid \tilde\psi\rangle=\hat\sigma_y\mid\psi^*\rangle,
\end{equation}
where $\hat\sigma_y$ is the Pauli y-matrix and $\mid\psi^*\rangle$ is the
complex conjugate of $\mid \psi\rangle$.
The entanglement, see  \cite{Wootters}, is defined as a function of concurrence, through the following equation 
\begin{equation}
E(\psi)=f(C(\psi)),
\end{equation}
where function $f(C(\psi))$ is given by
\begin{eqnarray}
f(C(\psi))&=&h(\frac{1+\sqrt{1-C(\psi)^2}}{2}),\\
h(x)&=&-x\log_2 x-(1-x)\log_2(1-x).
\end{eqnarray}
Function $f(C(\psi))$ increases monotonically from $0$ to $1$
as $C(\psi)$ ranges from $0$ to $1$. Therefore, the
concurrence can be considered as a measure of the entanglement. \\
The state describing our two-qubit system, written as a
superposition of the standard two-qubit computational basis
states, is given by
\begin{equation}\label{normal}
\mid\psi\rangle=c_1\mid00\rangle+c_2\mid01\rangle+c_3\mid10\rangle+c_4\mid11\rangle.
\end{equation}
Following eq. (\ref{spflip}), the spin-flip transformation over the state (\ref{normal}) gives 
\begin{equation}\label{sflip}
\mid\tilde{\psi}\rangle=-c_1^*\mid00\rangle+c_2^*\mid01\rangle+c_3^*\mid10\rangle-c_4^*\mid11\rangle.
\end{equation}
Finally, we obtain the concurrence of our system, see eq. (\ref{concurrence}), by performing the state product between states (\ref{normal}) and (\ref{sflip}). The normalized concurrence of the system is given by the following equation
\begin{equation}\label{concnorm}
C(\psi)=\frac{2\mid c_2^*c_3^*-c_1^*c_4^*\mid}{\mid c_1\mid^2+\mid
c_3\mid^2+\mid c_3\mid^2+\mid c_4\mid^2}.
\end{equation}
The result obtained in eq. (\ref{concnorm}) will be used to evaluate the degree of entanglement of our system during the
gate operation. When the concurrence related to a wave function reaches its maximum value,
the state is maximally entangled. Therefore, we require that the
concurrence of the wave function of the system, at the end of the gate operation, reaches a value next to its maximum.

\section{Phase Gate: Time Independent Case}\label{ind}
\subsection{Preliminary Setup}
\begin{figure}
\resizebox{0.80\columnwidth}{!}{

 \includegraphics{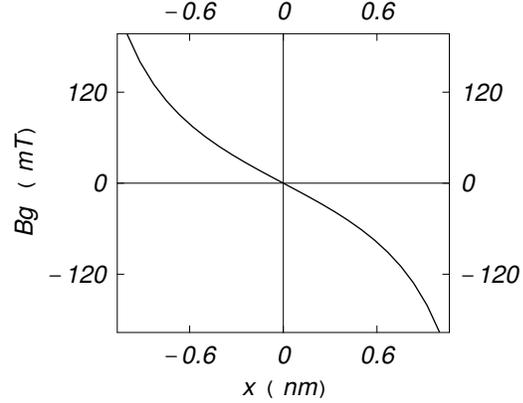}
      
}      
\caption{Magnetic field generated by two $1 \mu m$-radius wires at a
distance $d=1 \mu m$, which carry a current $I=0.6
A$. The two buckyballs are placed at a symmetrical distance $x$ with respect to the origin of the axes.}\label{gradfield}
\end{figure}

Our system is composed by two spins, which interact with a static
magnetic field. Applying a static magnetic field oriented in the
$z$ direction, for the Zeeman effect, we get the splitting of the
spin z component into the spin-up and spin-down components. The
energy difference between the two levels give the resonance
frequency of the particle. However, when we apply a static magnetic
field on the whole sample, all the particles will have the same
resonance frequency. To perform manipulations on each buckyball, we need to be able to distinguish each of them. This setup leads to the most relevant
experimental disadvantage for systems composed by arrays of
buckyballs, which is the difficulty in the individual addressing
of each qubit particle. This problem can be overcome with the use of external field gradients, which
can shift the electronic resonance frequency of the qubit-encoding
spins \cite{Harneit1,Suter}. Magnetic field gradients can be generated by considering wires
through which flows current. If we place two parallel wires
outside our two buckyball system, it is generated an additional magnetic
field in the space between the wires.
Following a paper by Groth {\it et al.} \cite{Groth}, with the help of atom chip technology, wires with a high current density can be built. The magnetic field amplitude generated by the two wires is given by
\begin{equation}
B_g=\frac{\mu_0}{2\pi}I(\frac{1}{x+\rho+d/2}+\frac{1}{x-\rho-d/2}),
\end{equation}
where $I$ is the current intensity, $d$ is the distance between the two wires, $\rho$ is the radius of each wire and $x$ is the distance of a buckyball with respect to the origin of the axes.
With the choice $I=0.6 A$, $d=1\mu m$ and $\rho=1 \mu m$, through a numerical computation, we obtain the magnetic field distribution shown in Fig. (\ref{gradfield}).
 We could not consider a current greater than $I=0.6A$ because the wires would face a too high heating process, and eventually they could be destroyed. On the other hand, we could not consider currents smaller than $10^{-1}A$, because the arising magnetic field gradient would be too small for each buckyball. In this case, the resonance frequencies related to the buckyballs would differ for only few $MHz$, which could be a too small gap to be realized by a frequency resonator.   
\subsection{Realization of the Phase Gate}
\begin{figure}
\resizebox{0.80\columnwidth}{!}{

 \includegraphics{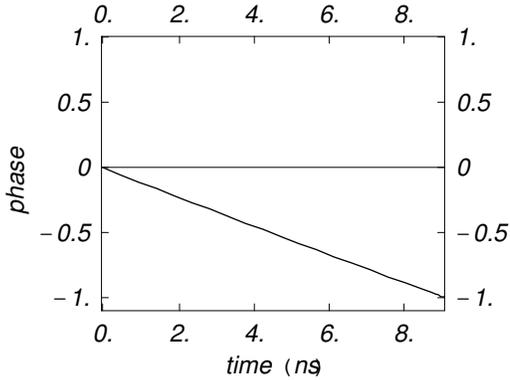}
      
}      
\caption{Time evolution of the phase $\vartheta$. The value
$\vartheta=-\pi$ is reached at the time $\tau\simeq 9.1\times
10^{-9}s$.}\label{tind}
\end{figure}

\begin{figure}
\resizebox{0.80\columnwidth}{!}{

 \includegraphics{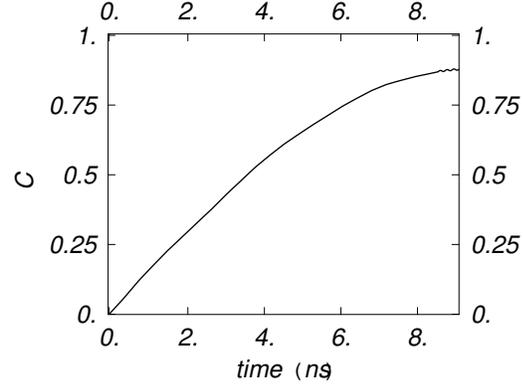}
      
}      
\caption{Time evolution of the concurrence,
$C(\psi)$.}\label{conctind}
\end{figure}

Choosing a static magnetic field in the z direction, the Hamiltonian of the system is given by the following equation
($\hbar=1$)
\begin{equation}\label{hamtind}
\begin{array}{lll}
H&=J_0\vec{\hat\sigma}_1\cdot\vec{\hat\sigma}_2+g(r)[\vec{\hat\sigma}_1\cdot
\vec{\hat\sigma}_2-3(\vec{\hat\sigma}_1\cdot\vec n)(\vec{\hat\sigma}_2\cdot\vec
n)]\\\label{hamtind1}
&-\mu_B[((B_{z_1}+B_{g_1})\hat\sigma_{z_1})\otimes
I_2\\
&+I_1\otimes((B_{z_2}+B_{g_2})\hat\sigma_{z_2})],
\end{array}
\end{equation}
In the previous equation, $J_0$ is the exchange spin-spin interaction coupling
constant, ${\hat\sigma}_1$ and ${\hat\sigma}_2$ are the Pauli spin matrices, $g(r)=\gamma_1\gamma_2\frac{\mu_0\mu_B^2}{8\pi r^3}$, where $\mu_0$ is the diamagnetic constant, $\mu_B$ is the Bohr magneton and $r$ is the distance between the two trapped atoms, $\vec n$ is the unit vector in the direction of the line which joins the centers of the two encased atoms,
$B_{z_1}=B_{z_2}$ is the static magnetic field in the $z$
direction, $B_{g_1}$ and $B_{g_2}$ are the additional magnetic
fields due to the field gradient. We make an assumption,
considering the trapped particles as electrons. Therefore the
gyromagnetic ratio $\gamma\simeq2$, and $g(r)=\frac{\mu_0\mu_B^2}{2\pi r^3}$.
 Through the study of fullerenes' spectra in  ESR (Electron Spin Resonance) experiments, and also
through theoretical studies, it has been shown \cite{Greer,Waiblinger,Harneit}, that the exchange
interaction is very small. Therefore, in eq. (\ref{hamtind}), we
can neglect the exchange term proportional to $J_0$, leaving the
spin dipole-dipole interaction as the leading term of the mutual
interaction between the two endohedrals. Choosing the direction of
vector $\vec n$ parallel to the $x$ axis, the dipole-dipole
interaction term is simplified as follows
\begin{equation}
\hat
D=g(r)(\hat\sigma_{z_1}\hat\sigma_{z_2}+\hat\sigma_{y_1}\hat\sigma_{y_2}-2\hat\sigma_{x_1}\hat\sigma_{x_2}).
\end{equation}
The Hamiltonian matrix form is given by the following matrix
\begin{equation}\label{matind}
\left(%
\begin{array}{cccc}
  g(r)+m_1 & 0 & 0 & -3g(r) \\
  0 & -g(r)+m_2 & -g(r) & 0 \\
  0 & -g(r) & -g(r)-m_2 & 0 \\
  -3g(r) & 0 & 0 & g(r)-m_1 \\
\end{array}%
\right),
\end{equation}
where
\begin{equation}
m_1=-\mu_B(B_{z_1}+B_{g_1}+B_{z_2}+B_{g_2})
\end{equation}
 and
\begin{equation}
m_2=-\mu_B(B_{z_1}+B_{g_1}-B_{z_2}-B_{g_2}),
\end{equation}
are the static magnetic field terms. Solving the 
\\Schr{\"o}dinger
equation
\begin{equation}\label{schrod}
\imath\frac{\partial}{\partial t}\mid \psi(t)\rangle=H\mid
\psi(t)\rangle,
\end{equation}
where the wave function is a superposition of the standard
two-qubit computational basis, given by equation (\ref{wave}), we
get the four differential equation system
\begin{eqnarray}\label{systemTindip1}
\dot c_1(t)&=&-\imath[(g(r)+m_1)c_1(t)-3g(r)c_4(t)];\\
\dot c_2(t)&=&-\imath[(-g(r)+m_2)c_2(t)-g(r)c_3(t)];\\
\dot c_3(t)&=&-\imath[
-g(r)c_2(t)+(-g(r)-m_2)c_3(t)];\\\label{systemTindip4} \dot
c_4(t)&=&-\imath[-3g(r) c_1(t)+(g(r)-m_1)c_4(t)],
\end{eqnarray}
which allows us to evaluate the phases acquired by each computational
basis state during the time evolution. Applying eq.
(\ref{vartheta}) to the present time evolved phases, we get the
desirable $\pi$-gate
\begin{eqnarray}\label{vartheta1}
\vartheta &=&Arg(c_1(t))-Arg(c_2(t))\\\nonumber
&-&Arg(c_3(t))+Arg(c_4(t))=\pm\pi,
\end{eqnarray}
where $Arg(c_i(t))$, $i=1,..,4$, which correspond to phases $\phi_{jl}$, $j,l=0,1$, in eq. (\ref{vartheta}), are the phases of  coefficients $c_i(t)$ of equation (\ref{wave}). 
We have numerically solved the differential equation system
(\ref{systemTindip1}-\ref{systemTindip4}), with the use of a
Mathematica programme. The numerical quantities used for the
numerical calculations are $r=1.14 nm$, $B_{z1}=B_{z2}=10\times
10^{-2}T$, $B_{g_1}=6.08 \times 10^{-5}T$ and $B_{g_2}=-6.08
\times 10^{-5}T$, which give the resonance frequencies
$\omega_1=1.7599\times10^{10}Hz$ and
$\omega_2=1.7577\times10^{10}Hz$. The time evolution of the phase
$\vartheta$ is shown in Fig. \ref{tind}. The gate time, which
corresponds to the case $\vartheta=-\pi$ is $\tau\simeq 9.1\times
10^{-9}s$. This result has been found for a chosen set of initial conditions $c_i(0)$, $i=1,..,4$. However, we did many trials for different numerical values of the set $c_i(0)$, $i=1,..,4$. In all these cases, phase $\theta$ shows a linear behavior and the resulting gate times are all in the same range, which is of the order of $10^{-8} s$. If the set of initial conditions is real, the starting value of phase $\theta$ is always equal to zero. If the set of initial conditions is complex, the starting value of $\theta$ is in the range $[-\pi,+\pi]$, but it can always be rescaled to zero. The numerical value of the distance between the two buckyballs, $r$, is a fixed value, which depends on the substrate where the buckyballs reside. The amplitude of the static magnetic field has been found by considering the allowed experimental limits for its realization. The chosen value for this amplitude has been found by checking the response of the system, i.e. the gate time, after some trials. Therefore, we can say that the phase gate time depends on the distance between the two buckyballs and on the amplitude of the static magnetic field, but it is independent of the choice of the initial values $c_i(0)$, $i=1,..4$.\\
 If we compare the gate-time, $\tau$, to the shortest
decoherence time, $T_2\simeq20\mu m$, we can deduce that it will
be theoretically possible to realize about thousands gate
operations before the system relaxes.
To know the fidelity of the gate and the reliability of the
results, we need to evaluate the concurrence during the time
evolution. With the use of a Mathematica programme we have plotted
the time evolution of the concurrence, equation (\ref{concnorm}),
from $t=0s$ to the gate time $t=\tau$, see Fig. \ref{conctind}.
Analyzing picture (\ref{conctind}), we can see that the
concurrence shows a smooth behavior. It monotonically ranges from
zero and its maximum is reached at time $t=\tau$, with the
respective value $C(\psi(\tau))=0.88$. Even if the maximum
concurrence does not coincides with the ideal value $1$, it is
near to this value and the system shows an acceptable degree of
entanglement. It is convenient to investigate other system
configurations, in order to check if it is possible to improve the
concurrence. In the next Section we will analyze the case of an
additional magnetic field, oscillating in time in the x-y plane.
\section{Phase gate: Time Dependent Case.}\label{dip}
\begin{figure}
\resizebox{0.80\columnwidth}{!}{

 \includegraphics{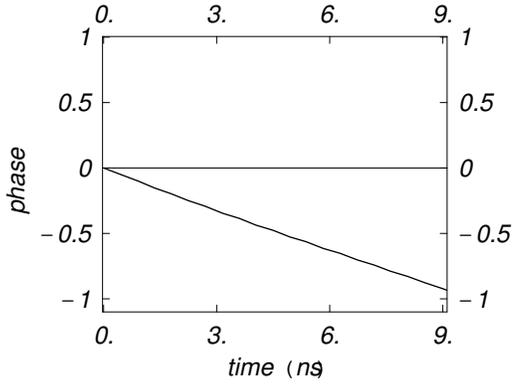}
      
}      
\caption{Time evolution of the phase $\vartheta(t)$, with the respective gate time $\tau\simeq9.8\times 10^{-9}s$.}\label{tdip}
\end{figure}

\begin{figure}
\resizebox{0.80\columnwidth}{!}{

 \includegraphics{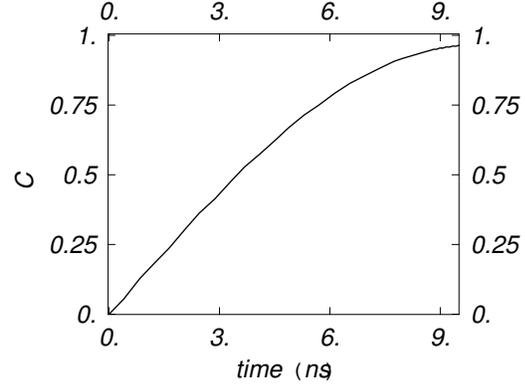}
      
}      
\caption{Time evolution of the concurrence,
$C(\psi)$.}\label{conctdip}
\end{figure}

In this Section, we apply to our system an additional time
dependent magnetic field. To induce the transitions between the
two Zeeman energy levels, we need to apply an oscillating magnetic
field in the $x-y$ plane with angular frequency, $\omega$, equal
to the spin resonance frequency. 
In the case of a transverse linear oscillating
magnetic field, the total applied magnetic field is given by
\begin{equation}
\vec B(t)=(B_l\cos \omega t,B_l\cos \omega t,(B_z+B_g)).
\end{equation}
The Hamiltonian of the system reads
\begin{equation}
\begin{array}{lll}
H&=g(r)(\sigma_{z_1}\sigma_{z_2}+\sigma_{y_1}\sigma_{y_2}-2\sigma_{x_1}\sigma_{x_2})\\
&-\mu_B (B_{z_1}+B_{g_1})\sigma_{z_1}\otimes I_2\\
&- \mu_B(
B_{z_2}+B_{g_2})I_1\otimes
\sigma_{z_2}\\
&-\mu_B B_{l_1}(\sigma_{x_1}\cos \omega_1 t+\sigma_{y_1}\cos
\omega_1t)\otimes I_2\nonumber\\
&+I_1\otimes(-\mu_B B_{l_2}(\sigma_{x_2}\cos \omega_2
t+\sigma_{y_2}\cos \omega_2t)).
\end{array}
\end{equation}
Like in the time independent case, solving the Schr{\"o}dinger
equation, we get a four differential equation system,
whose solution give the time evolution of the phase for each
computational basis state. Arranging the phases as prescribed in
equation (\ref{vartheta}), we have obtained the $\pi$-gate. In the
numerical computation we have used the additional quantity
$B_{l_1}=B_{l_2}=5\times 10^{-4}T$.
shown in Fig. \ref{tdip},
 and the numerical value of the gate time is $\tau\simeq
9.8\times 10^{-9}s$. Also in this case, comparing the gate time,
$\tau$, to the decoherence time $T_2$, we observe that it will be
possible to perform about thousands gate operations before the
system relaxes. The relevant result in the treatment of the time
dependent case is the concurrence. In Fig. \ref{conctdip},
 it is represented the time evolution of the concurrence,
$C(\psi(t))$, which has been numerically evaluated with a
Mathematica programme. It shows a monotonic behavior and its
maximum, evaluated at time $t=\tau$, corresponds to
$C(\psi(\tau))=0.96$. Therefore, an additional
linearly polarized oscillating field in the $x-y$ plane allows the
system to be characterized by a better concurrence degree.
\section{Conclusions}
To model quantum gates we considered a system composed by two
endohedral fullerene molecules, subjected to external magnetic
fields. We assume that each molecule may be treated as a
$\frac{1}{2}$-spin particle, where the spin is associated to the
encapsulated atoms. In the magnetic field the spin degeneracy of
the spin up and down components is lifted and it arises the Zeeman
splitting. As the result, there two two-level system are arising.
Each of these two-level systems corresponds to a single qubit. If
the applied static magnetic field to the whole sample is
homogeneous, each of these qubits will be characterized by the
same resonance frequency. This leads to the difficulty in the
individual addressing of each single qubit. To overcome this
problem, we have to apply inhomogeneous magnetic fields. in this
paper we have used a magnetic field generated by two metallic
wires. Each wire is carrying a current, therefore the magnetic
field is decreasing with the distance from a wire. In the proposed
configuration of two parallel wires, there arises a gradient of
the magnetic field when we are moving from a wire to the other
one. If we place two buckyballs in the space between these two
wires, they will be subjected to the gradient of this field, and
therefore the associated resonance frequencies of the related
two-level system are different. In this paper we have performed a
quantum $\pi$-phase gate. To realize this particular quantum gate
we have estimated the phase of each computational basis state, see
equation (\ref{vartheta}). The leading mutual interaction between
the two qubits is the spin dipole-dipole interaction. First we
studied the time evolution of our system taking into account this
mutual interaction between the qubits and considering the qubits
subjected to static magnetic fields only. Then we applied to the
system also time dependent magnetic fields. The wave function of
the system is given by the superposition of the four computational
basis states, see equation (\ref{wave}). The time evolution of the
coefficients of each computational basis state is determined via
the solution of the Schr{\"o}dinger equation. With the use of
these coefficients and of equation (\ref{vartheta1}), we can
evaluate the operational gate time for the $\pi$-phase gate. Its
numerical value is $\tau\simeq9.1\times 10^{-9}s$ for the time
independent case, and $\tau\simeq9.8\times 10^{-9}s$ for the time
dependent one. Comparing the gate time, $\tau$, to the shortest
relaxation time, $T_2$, we have observed that in both cases it
will be possible to perform about thousands quantum gate
operations before the system decoheres. This is our main result.
As far as we are aware, this result indicates that our system
could be the most favorable for the realization of a quantum gate.
Obviously, for realistic models of quantum computers, the ratio of
the decoherence time and the operational time must be very large,
otherwise the system relaxes before the completing of the quantum
computation. The goal of any quantum computational proposal is the
entanglement of the state of the system under consideration. At
this purpose, we have studied the concurrence, see Sec.
\ref{concurr}. The concurrence gives information about the
entanglement of the state, therefore it is related to the
reliability of the gate operation. A maximally entangled state is
left unchanged under a spin-flip operation and its concurrence is
maximum. In our system, at the end of the gate operation, the
value of the concurrence is $C\simeq0.88$ in the time independent
case, and $C\simeq0.96$ in the time dependent one. Both values are
acceptable because they are both related to  a very good degree of
entanglement for the state describing our system. We can conclude
that the best configuration for our system is the time dependent
one. It is characterized by a very small operational time, in
comparison to the relaxation
times, and by the best concurrence.\\
Many features claim the buckyball systems as good candidates for performing 
quantum gates. Not only they are characterized by very long
decoherence times, but also they can be maneuvered very easily.
This feature allows the realization of experimental quantum
devices, which form scalable architectures. For example,
buckyballs can be embedded in silicon surfaces or arranged in arrays
encased in a nanotube (peapod). Moreover, in such systems we
suppose that the value of the relaxation time $T_2$, due to random
spin dipole-dipole interactions, could be reduced.

\end{document}